# Structure and stability of carbon nitrides: ring opening induced photoelectrochemical degradation

Florentina Iuliana Maxim[a], Eugenia Tanasa[b], Bogdan Mitrea[a], Cornelia Diac[a], Tomas Skala[c], Liviu Cristian Tanase[d], Adrian Ciocanea[e], Stefan Antohe[f], Eugeniu Vasile[b], Serban N. Stamatin[a,f*]

[a] 3Nano-SAE Research Centre, University of Bucharest, Atomistilor 405, 077125, Magurele, Ilfov, Romania

[b] Department of Oxide Materials and Nanomaterials, Faculty of Applied Chemistry and Material Science, University POLITEHNICA of Bucharest, Bucharest 060042, Romania

[c] Department of Surface and Plasma Science, Charles University, V Holešovičkách 2, 18000 Prague, Czech Republic

[d] National Institute of Materials Physics, Atomistilor 405A, 077125, Magurele, Ilfov, Romania

[e] Hydraulics and Environmental Engineering Department, Power Engineering Faculty, Hydraulics, University POLITEHNICA of Bucharest, 060042 Bucharest, Romania

[f] Faculty of Physics, University of Bucharest, Atomistilor 405, 077125, Magurele, Ilfov, Romania

[*]corresponding author telephone no.: +40214574838 and e-mail address: serban@3nanosae.org

**Abstract**

A constant increase in the need of clean energy demands more innovation from the research community. Active and stable materials that can make use of the solar radiation to promote different reactions are the cornerstone of emerging technologies. Polymeric carbon nitrides that harvest solar radiation to drive electrochemical reactions are considered solid candidates. In this respect, polymeric carbon nitrides were prepared by the thermal polycondensation of melamine. As-obtained materials were characterized by synchrotron radiation and lab-based techniques. The electron band structure was fully characterized by a combined electrochemical – optoelectronic study. Electron microscopy studies before and after the photoelectrochemical experiments showed morphological and structure degradation. The work at-hand concludes that polymeric carbon nitrides are prone to photoelectrochemical degradation at high overpotentials.

## 1. Introduction

Sun driven electrochemical reactions, such as (1) photocatalytic environmental remediation, (2) photoelectrochemical water splitting to generate hydrogen fuel and (3) photoelectrochemical $CO_2$ reduction to produce chemical feedstocks, will have the utmost importance in securing sustainability. The electron-hole pair generated by illuminated semiconductors are the key in unlocking the full potential of photoelectrochemistry. More than 50 years have passed since the seminal paper by Fujishima and Honda[1] that lead to the intense research in finding materials with high photoelectrochemical activity, such as La doped $NaTaO_3$[2]. Further on, the attention was focused on finding cheap and abundant materials such as polymeric carbon nitride[3] which has been in the spotlight for more than a decade.

Polymeric carbon nitrides are obtained from the thermal polycondensation of nitrogen rich precursors, such as melamine, cyanamide and other similar compounds. The structure and surface chemistry of carbon nitrides obtained by thermal polycondensation has been the subject of intense debate in the literature[4–11]. It was initially believed that the product of the polycondensation is the highly stable carbon nitride (i.e. $C_3N_4$). Ideal $C_3N_4$ consists of a carbon-nitrogen network which is challenging to obtain by heat treatment due to the presence of hydrogen and other impurities in the precursors and air. Indeed, it was shown that $C_3N_4$ contains hydrogen below 3% and its structure resembles that of melon (i.e. $C_6N_9H_3$)[10,11]. The term graphitic carbon nitride (i.e. g-$C_3N_4$) is still used in the literature albeit the structure and surface chemistry are fundamentally different. Carbon nitride is inherently more stable than melon which made stability issues improbable and therefore such studies are not present in the literature. The term g-CN will be used instead of g-$C_3N_4$ in the work at-hand to underline the fundamental difference between two classes of materials.

The photoelectrochemical activity of pristine carbon nitride is not that high but can be further tuned by doping, annealing, increased surface area, chemical modifications etc[12–15]. The nature of active sites and fundamental reaction mechanisms are very scarce in the literature, despite the large body of work available in the literature on polymeric carbon nitrides. Stability investigations are paramount in establishing active sites and reaction mechanisms, as it can bring to light reactant species. It was shown that the nitrogen concentration decreases during photocatalytic experiments which indicates that polymeric carbon nitrides are not stable under photocatalytic conditions, that is powder dispersed in a liquid[16]. In contrast to photocatalysis, photoelectrocatalytic conditions means that a controlled potential is applied on a photoelectrode. Polymeric carbon nitride photoelectrodes are obtained by spray deposition on a transparent fluorine doped tin oxide (FTO) glass. It is obvious that at large potentials the interface between the carbon nitride layer and FTO will be severely affected, especially in gas evolving experiments (e.g. hydrogen evolution reaction or $CO_2$ reduction).

Polymeric carbon nitrides were initially tested towards hydrogen evolution[3], a reaction that does not involve complex surface intermediates. The interest has shifted towards more complex reactions, such as $CO_2$ reduction[15]. Electrochemical $CO_2$ reduction is a considerably more complex reaction which involves many steps and adsorbed intermediates even on metallic surfaces[17]. Electrochemical reduction reactions at neutral pH usually start with the formation of radicals, such as superoxide radical or $CO_2$ anion radical[18–20]. It is expected for a defective carbon-based structure to be prone to corrosion that can alter the initial morphology and surface chemistry which leads to deactivation. In this respect, carbon nitride stability investigations are vital in reaching beyond the state-of-the-art.

Herein, two common polymeric carbon nitrides, yellow and red (inset **Figure 1**), were prepared by thermal polycondensation route. Subsequent annealing of the polymeric carbon nitride resulted in what is known as *red* or *amorphous carbon* nitride[21,22]. Synchrotron radiation was used to carry out photoelectron emission and absorption spectroscopy which resulted in an in-depth understanding of the structure and surface chemistry. The morphology was assessed by high resolution electron microscopy. The electronic band structure was determined based on the optoelectronic characterization. Electron microscopy investigations before and after photoelectrochemical experiments, also known as post-mortem studies, showed extensive structural degradation which was further confirmed by FT-IR.

## 2. Experimental section

*2.1. Reagents*

NaHCO$_3$ (S.C. Herodio Crafts S.R.L., Romania), KCl (Chimactiv S.R.L., Romania), K$_3$[Fe(CN)$_6$] (Synteza, Poland), melamine (Sigma Aldrich, USA) and ethanol (S.C. Tunic Prod S.R.L., Romania) were used without further purification. Glass slides coated with a fluorine doped oxide (FTO) with a surface resistivity of 7 Ω cm$^{-2}$ was supplied by Sigma-Aldrich (#735167-1EA, St. Louis, Missouri, USA), then cut into 3 cm x 3 cm pieces. A Milli-Q DirectQSystem (Burlington, MA, USA) was used for obtaining ultrapure water (> 18.2 MΩ cm).

## 2.2. g-CN and g-CN-HT synthesis

Metal free carbonitride (g-CN) was prepared by thermal polycondensation of melamine at 550°C for 4 hours. The precursor was placed in a covered alumina crucible, in air, and heated to 550°C with a 5°C min$^{-1}$ heating rate. After the synthesis, the crucible was freely cooled to room temperature. Annealed graphitic carbon nitride (g-CN-HT) was obtained by annealing g-CN for 4 hours at 675°C under Ar atmosphere. The final products had a yellow (g-CN) and a red (g-CN-HT) color.

## 2.3. Structural and morphological characterization

Photoemission and X-ray absorption experiments were carried out at the Materials Science Beamline at Elettra Synchrotron facility in Trieste, Italy. It is a bending-magnet beamline with a tuning range 22-1000 eV. An ultra-high vacuum experimental chamber (base pressure below 2×10^-10 mbar), a hemispherical electron analyzer (Specs Phoibos 150), sample manipulator and load lock were used to conduct the experiments. Photoemission measurements were performed at normal emission (60° incidence) geometry with photon energies 700 eV (core levels O 1s, N 1s, C 1s; total resolution 0.5-0.7 eV). Changes in the core level spectra were caused by the synchrotron radiation during photoelectron emission experiments. XPS measurements are shown only with the Al Kα (1487 eV, total resolution 1eV). The photocurrent generated by a gold mesh was used to normalize the spectra. X-ray absorption (NEXAFS) measurements were performed in the Auger-electron yield mode near the C and N K-edges. The resolution was 0.3-0.4 eV. The energy scale was calibrated using data measured on an Ar-sputtered gold foil. Reference data from this sample were also used for intensity normalization of the NEXAFS spectra.

Morphological characterization was done on a scanning electron microscope (SEM) and transmission electron microscope (TEM) operated at 30kV and 300 keV, respectively. The SEM (Quanta Inspect F) and TEM (Tecnai $G^2$ F30 S-Twin) were both manufactured by FEI-Philips (Hillsboro, OR, USA). A Panalytical X'PERT PRO (Almelo, The Netherlands) was used to carry out powder X-ray diffraction (Cu Kα = 1.54 Å). UV-Vis spectroscopy was done by means of a Jasco spectrophotometer (V550, Tokyo, Japan).

*2.4 Electrochemical measurements*

Inks, containing 50 mg carbon nitride powder in 5 mL ethanol, were homogenized for 15 min using an ultrasonic processor (Ultrasonics FS-250N, MXBAOHENG, Zhejiang, China). The inks were sprayed on heated (80°C) FTO glass plates (working electrode support). Polymeric carbon nitride films were obtained by means of an air brush operated at 2 bar. The air brush was manufactured by HSENG, China, model AF186 Mini Air Compressor.

Carbon dioxide reduction reaction and Mott – Schottky measurements were measured by means of a Voltalab electrochemical workstation (Voltalab, PGZ 301). The photo-electrochemical cell (PECC-2) was manufacture by ZAHNER in Germany. The photo-electrochemical cell was equipped with a transparent working electrode with an 18 mm optical window diameter. A mercury lamp (PS-2, China) was used for front illumination. FTO glass deposited with various g-CN served as a working electrode. A platinum wire was used as a counter electrode. The reference electrode, Ag/AgCl, potential was measured against the standard hydrogen electrode[23,24]. The reference electrode potential was measured 5 times with an average of 0.201 V +/- 0.001 V (95% confidence interval).

Carbon dioxide reduction reaction under visible light was tested on the g-CN and g-CN-HT. $CO_2$ was purged in the PEC cell filled with 0.1 M $NaHCO_3$ for twenty minutes before the electrochemical measurements. The reduction photocurrents of each sample were measured under different applied potentials (– 0.3 V, – 0.4 V, – 0.5 V, – 0.6 V, – 0.7 V). Each measurement consisted of 4 cycles and each illumination cycle had 30 seconds dark and 30 seconds illuminated.

Mott-Schottky measurements were based on the determination of the inverse squared capacitance by electrochemical impedance spectroscopy in 0.1 M KCl with 1 mM $K_3Fe(CN)_6$. Dynamic electrochemical impedance spectroscopy was performed between –1 V and –2 V range vs. Ag/AgCl at a fixed frequency of 100 kHz, 10 mV amplitude and a potential step of 0.025 V.

## 3. Results and discussion

*3.1. Material synthesis*

The solid-state pyrolysis of melamine was used to obtain graphitic carbon nitride (g-CN), further details can be found in the Experimental section. Elemental analysis showed that g-CN consisted of 35.5% carbon, 62.6% nitrogen and 1.1% hydrogen. The as-obtained g-CN was placed in a horizontal tube furnace at 675 ºC with a constant Ar flow which resulted in a red powder, hereinafter g-CN-HT (Figure 1). Elemental analysis showed a slightly different C/N/H ratio for g-CN-HT, that is 35.8% carbon, 62.0% nitrogen and 0.7% hydrogen. The carbon to nitrogen ratio (C/N) was determined to be 0.57 and 0.58 for g-CN and g-CN-HT, respectively. The C/N theoretical value for melon and g-C$_3$N$_4$ is 0.67 and 0.75, respectively, which are significantly higher than the samples under investigation. Elemental analysis showed that the samples are similar in bulk composition, although considerably different in structure, morphology and surface chemistry (*vide infra*).

*3.2. Structure and morphology*

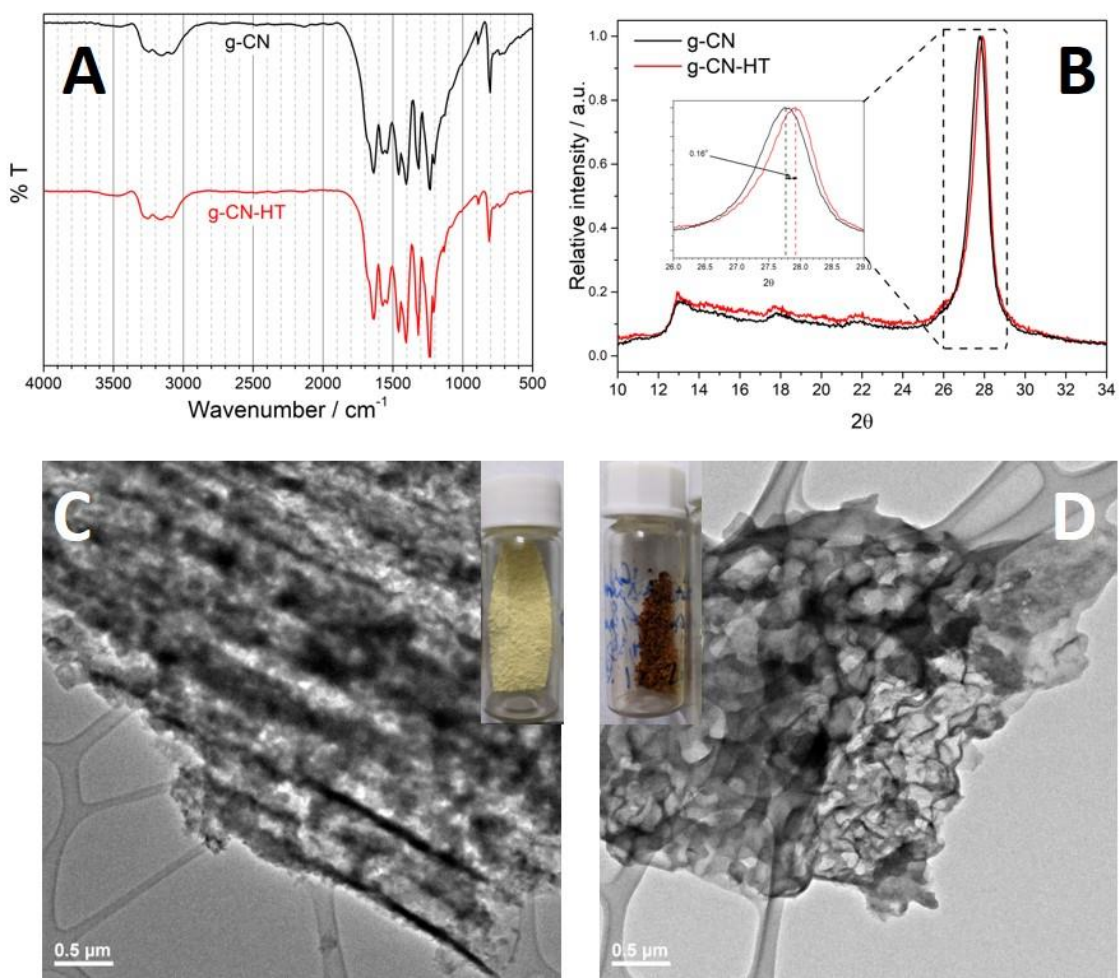

**Figure 1** Structure and morphology of g-CN and g-CN-HT. Fourier-transform infrared spectroscopy of g-CN, black line, and g-CN-HT, red line (A); X-ray diffractograms of g-CN, black line, and g-CN-HT, red line (B); transmission electron microscopy of g-CN (C) and g-CN-HT (D)

The structure of polymeric carbon nitrides can be qualitatively assessed by Fourier transform – infrared (FT-IR)[4,25–27] which is shown in Figure 1A. The peaks were identified based on the first modern synthesis work of polymeric carbon nitrides authored by Komatsu[25]. Heptazine (i.e. cyameluric nucleus) can be identified in the FT-IR by its specific absorption band at 810 cm$^{-1}$ which was visible for both samples. The region between 1200-1700 cm$^{-1}$ has 6 main peaks corresponding to different bonds in the heterocycle. The peak at 1233 cm$^{-1}$ can be attributed to the nitrogen bonded to 3 carbons in the trigonal units, as in C-N(-C)-C. The bands at 1457 and 1571 cm$^{-1}$ correspond to the ring vibrations. The peaks at 1403 and 1637 cm$^{-1}$ correspond to the δ(NH) and δ(NH$_2$), respectively. The last region defined between 2800-3400 cm$^{-1}$ can be assigned to ν(NH).

X-ray diffraction (XRD) was used to probe the long-range order of the materials under investigation (Figure 1B). Four peaks were observed in the region between 2θ = 12° and 2θ = 32°. The peak at 2θ = 27.76° has the highest intensity followed by the peak at 2θ = 13.12°. Previous XRD studies showed that the polymeric carbon nitride resulted from solid-state pyrolysis of melamine is in fact melon, although the presence of heptazine-based g-C$_3$N$_4$ cannot be fully excluded on the sole basis of XRD[6,10,11]. The XRD profile of g-CN indicated an orthorhombic geometry with main diffractions at 2θ = 27.76° and 2θ = 13.12° which were assigned to (002) and (210), respectively. Available structural models showed that the separation between parallel melem units gives rise to (210) while (002) is attributed to the separation between graphitic sheets[6,11]. An interlayer distance of 3.22 Å was obtained for g-CN based on the (002) Bragg peak which is similar to other reported values[10,11]. The interlayer distance for g-CN-HT was determined to be 3.19 Å, resulted from the 0.16° increase in 2θ (inset of Figure 1B). The intensity ratio between (210) and (002), $I_{210}/I_{002}$, is another important feature in the XRD profiles of carbon nitrides[6]. The calculated $I_{210}/I_{002}$ was 0.17 and 0.20 for g-CN and g-CN-HT, respectively. The $I_{210}/I_{002}$ values are similar to the theoretical value, that is 0.19.

The morphology of g-CN and g-CN-HT (Figure 1C-D and Figure S1, S2) was assessed by electron microscopy. A thorough SEM and TEM investigation exposed the tube-like morphology of g-CN (Figure S1A-B and black arrows in S2A-B, Supplementary Materials). Representative SEM images of g-CN showed that tubular structures were bundled into larger formations (Figure S2A, Supplementary Materials). A fractured piece showed cracked tubes which were well-aligned and tightly packed together (Figure S1B, Supplementary Materials). The length of the tubes was determined to be around 2-5 μm with a tube diameter below 300 nm (Figure S2A, Supplementary Materials). The tubes were densely packed into ordered structures mixed with an amorphous component. Individual tubes were observed only in the proximity of an amorphous fraction (black arrows in Figure S1A-B, Supplementary Materials).

The morphology of g-CN-HT is considerably different than that of g-CN. Annealing g-CN degrades the tubes into thin-sheets lumped into larger fractions (Figure S1C-D, Supplementary Materials). TEM investigations of g-CN-HT (Figure 1D) showed an amorphous structure without any ordering which is further supported by the SEM (Figure S2C-D, Supplementary Material). Further TEM investigations revealed that the tubes present in g-CN have completely lost their tubular geometry (red arrows in Figure S1C-D, Supplementary Material). Similar morphological features were encountered for amorphous and red carbon nitrides [21,22].

*3.3. Surface chemistry characterization by synchrotron radiation*

Synchrotron-based near edge X-ray absorption fine structure (NEXAFS) was used to characterize the chemistry of the materials under investigation (Figure S3 see Supplementary Material). The reader should bear in mind that NEXAFS on carbon nitrides is still a topic of hot-debate in the literature[27–30]. NEXAFS was used in this work as a fingerprinting technique to map the nitrogen chemistry to deconvolute the N 1s XPS (*vide-infra*). The NEXAFS for g-CN and g-CN-HT showed a similar profile to previous reports[27–30]. The highest intensity peak at 399.3 eV (A) is followed by 2 smaller features around 401 eV (B and C) and the second most intense peak at 402.1 eV (D). The largest contribution to the NEXAFS (A) was ascribed to the $sp^2$ hybridized nitrogen, as in C=N-C in the heptazine unit[29], which is also the most abundant nitrogen bonding in the structure (*vide-infra*). Peak B was assigned to the central tertiary nitrogen, as in $(C)_3$-N in heptazine. Peak C was assigned to terminal nitrogen as in C-$NH_2$ [29]. It is generally accepted that the nitrogen bridging heptazine units, as in $(C)_3$-N, has a resonance around 402 eV[27], however, it was shown recently that it overlaps with the $sp^2$ hybridized nitrogen[29,30]. In consequence peak D was assigned to the $sp^2$ hybridized nitrogen[29]. Peak assignments in NEXAFS (Figure S3, Supplementary Material) was consistent with the most recent equivalent core-hole time-dependent density functional theory modelling[29] which showed the contribution of four different nitrogen chemistry.

FT-IR and XRD showed that g-CN and g-CN-HT possess a similar structure while the SEM and TEM showed that their morphology differs significantly. Core-level X-ray photoelectron spectroscopy (XPS) was used to probe the surface chemistry of g-CN and g-CN-HT in order to achieve an improved understanding of the materials (**Figure 2**). Wide scan XPS showed that the samples contained only C, N, O and Si (Figure S4, Supplementary Material). Samples prepared for XPS were deposited on Si which explains the presence of Si in the wide scan.

N 1s core-level spectrum of g-CN is presented in Figure 2A. The deconvolution of the N 1s XPS into 4 main contributions was in line with the NEXAFS findings and previous literature reports[10]: (1) the $sp^2$ nitrogen located at 399.0 ± 0.1 eV; (2) the nitrogen in the amino group, as in C-$NH_2$, located at 399.5 ± 0.1 eV; (3) the nitrogen linking two melem units, hereinafter as $(C)_2$-NH, located at 400.5 ± 0.1 eV and (4) the central nitrogen bonded to three carbons in the central heptazine ring, hereinafter $(C)_3$-N, located at 401.5 ± 0.1 eV. The component at binding energy larger than 403.5 eV was attributed to a shake-off type of change in Coulombic potential, generally known as XPS satellite peaks. The ratios of the area intensity for each peak relative to the total intensity are presented in Table 1. The previous elemental analysis and XRD investigations suggested that melon is a more representative structure for g-CN and g-CN-HT, rather than g-$C_3N_4$. Melon's ideal structure consists of C=N-C, C-$NH_2$, $(C)_2$-NH and $(C)_3$-N in 67:11:11:11 proportions[10] which was found to be very similar for g-CN, that is 69.5:9.6:11.1:9.8 (Table 1). This underlines g-CN's similarity to melon rather than g-$C_3N_4$, although there were minor deviations from the ideal structure of melon. A similar approach was pursued for the surface chemistry of g-CN-HT which was determined to be: 66.2:12.2:13.1:8.5 (Table 1). A decrease in the surface concentration of C=N-C and $(C)_3$-N was observed between g-CN and g-CN-HT (Table 1) which corresponds to the breakage of the s-triazine and heptazine units. The increase in the surface concentration of $(C)_2$-NH observed for g-CN-HT (Table 1) gave more substance to the disruption of the heptazine-based g-$C_3N_4$ structure. A similar rationale was used for the increase in the surface concentration of C-$NH_2$ which indicates that the s-triazine units were undergoing structural change.

The focus was then turned to carbon's core-level XPS spectra (Figure 2B and D). NEXAFS was used for fingerprinting which showed a main peak at 287.9 eV ($C_2$) with a smaller feature at 287 eV ($C_1$; Figure S3, Supplementary Material). The main peak, $C_2$, was assigned to the absorbing carbon atom in N=C-$(N)_2$ while distortions of the local structure was found to lead to a feature at smaller binding energy[29]. The XPS core-level spectra of C 1s for g-CN and g-CN-HT (Figure 2B and D) showed a main peak at 288.6 eV which was assigned to $sp^2$ carbons of the heptazine rings, as in N=C-(N)$_2$ as it was previously shown[10]. The second most intense peak at approx. 285 eV was assigned to adventitious carbon, as in C-C. Hydrocarbon impurities easily adsorb on the sample support or even on the sample which is usually associated with adventitious carbon. An indium tin oxide foil was used as a sample support. XPS characteristic peak of In 3d is located at a binding energy of 444 eV which is clearly not visible in the wide spectrum (Figure S4, Supplementary Material). The XPS probing depth (i.e. photoelectron inelastic mean-free-path) is less than 5 nm and the XPS samples were prepared by spray coating which generates micrometer thick films. Therefore, it is more likely for the origin of adventitious carbon to be the sample rather than the support. The peak labeled as impurities in Figure 2B and D was located at approx. 286-287 eV which is an area specific for oxygen moieties on carbon surfaces[24,31–33] which can be related to oxidized hydrocarbons.

The carbon to nitrogen ratio (C/N) on the surface was determined by dividing the N=C-(N)$_2$ area in C 1s to the nitrogen area, excluding satellites (Table 1). An ideal g-C$_3$N$_4$ should have a 0.75 C/N ratio while the same ratio for melon should be 0.67[10,11]. Both samples under investigation showed a C/N ratio smaller than the ideal g-C$_3$N$_4$ even in the first few nm. The bulk C/N ratio in both samples was below 0.6 (*vide supra*). The results point to a chemical structure and composition similar to that of melon at the nanoscale, albeit significantly different in bulk.

**Table 1.** Surface chemistry determined as% from the XPS core-level spectra of N 1s and C 1s

|  | C/N | N 1s / % | | | | C 1s / % | | |
|---|---|---|---|---|---|---|---|---|
|  |  | C=N-C | C-NH$_2$ | (C)$_2$-NH | (C)$_3$-N | C-C | imp | N=C-(N)$_2$ |
| g-CN | 0.68 | 69.5 | 9.6 | 11.1 | 9.8 | 14.3 | 1.3 | 84.3 |
| g-CN-HT | 0.72 | 66.2 | 12.2 | 13.1 | 8.5 | 20.4 | 1.5 | 78.0 |

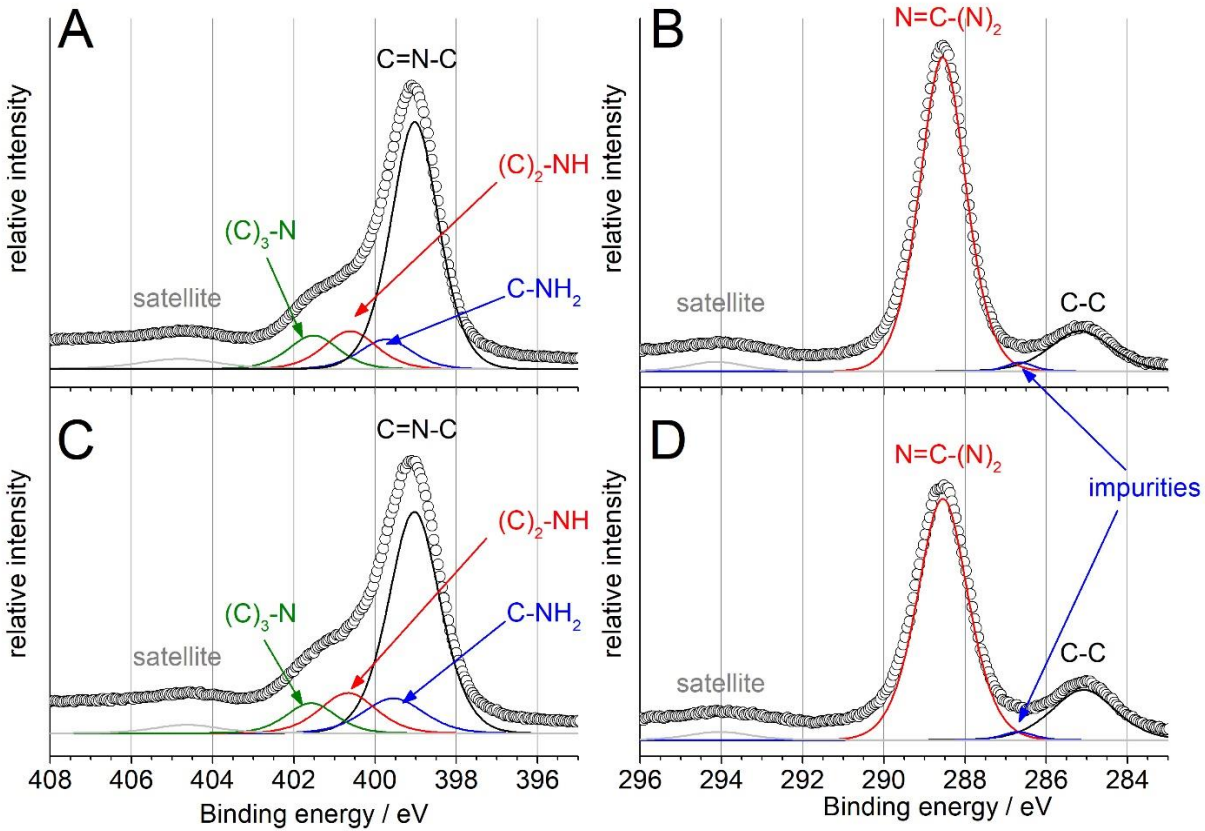

**Figure 2** X-ray photoelectron spectroscopy (XPS) investigation on g-CN (A-B) and g-CN-HT (C-D). N 1s core-level spectra for g-CN (A) and g-CN-HT (C). C 1s core-level spectra for g-CN (B) and g-CN-HT (D).

*3.4 Photoelectrochemical characterization*

*Electronic band structure*

The UV-Vis spectra (Figure 3A) of both samples exhibited the typical carbon nitride profile with a semiconductor bandgap onset. g-CN showed a main peak at 370 nm consistent with the yellow color. The π-π* electron transition gives rise to the peak at 370 nm (Figure 3A). The absorption of g-CN-HT was at 520 nm which is also known as a red-shift. The n-π* electron transitions involving lone pairs on the N atoms situated at the edge is believed to be the source of the 520 nm peak[15,21,22]. The reader should bear in mind that n-π* electron transitions are forbidden in ideal structures based on s-triazine repeating units[34,35]. A Tauc plot was constructed to obtain the optical bandgap (inset Figure 3A). The *x-axis* intercept of the linear region in the Tauc plot showed an optical bandgap of 2.60 eV and 2.21 eV for g-CN and g-CN-HT, respectively. Similar bandgap values were obtained for pristine and heat-treated carbon nitrides[12,15]. Carbon nitrides with a narrower bandgap have been obtained before by heat treatment or alkaline earth metal doping[15,21,22,35]. Annealing is a well-known technique to improve the crystallinity without disrupting the 3D long-range order, however, this is not the case for nitrogenated carbons[19,20].

The flat band potential, $V_{fb}$, can be obtained from the electrochemical impedance spectroscopy measured in dark[36]. The specific capacitance is related to the electrode applied potential by the Mott-Shottky equation (see Supplementary Material). The values reported in this work are shifted by -0.61 V (-0.2 V for the Ag/AgCl reference potential and -0.41 V for the potential – pH correction) from the values reported against the normal hydrogen electrode (NHE) at pH = 0[12]. Typical values for electrochemical conduction band of carbon nitrides were found to span between -1 and -1.2 V vs. NHE which is equivalent to -1.61 and -1.81 V vs. Ag/AgCl[12] (taking into account the corrections mentioned above).

Figure 3C shows that the $CO_2$ reduction reaction ($CO_2$RR) photocurrent density is smaller than 0.5 µA cm$^{-2}$ for both samples. The maximum photocurrent density for g-CN and g-CN-HT was found at -0.4 V vs. Ag/AgCl and -0.7 V vs. Ag/AgCl, respectively. There are many factors at play in generating the photocurrent, such as photoelectrode loading, deposition method and so forth. Nevertheless, photocurrent densities ranging from 0.1 µA cm$^{-2}$ to 10 µA cm$^{-2}$ are common in the literature[15,37–39]. The trend in photocurrent density is steadily increasing with increasing potential for g-CN-HT. A decreasing trend in photocurrent density was observed for g-CN after reaching the maximum at -0.4 V vs. Ag/AgCl. Figure 3D shows the electronic band structure for g-CN and g-CN-HT based on the bandgap determined from the Tauc plot (inset Figure 3A) and Mott-Schottky plot (Figure 3B). The right-hand axis in Figure 3D shows the most probable electrochemical reactions to take place at the photoelectrodes presented in Figure 3C. All the reactions have a standard potential between -0.3 and -0.8 V vs. Ag/AgCl. Other reactions were excluded to simplify the image considering that product formation and selectivity falls beyond the scope of this manuscript, that is photoelectrochemical corrosion. The conversion efficiency and product selectivity during the photoelectrochemical CO2RR over carbon nitrides has been extensively studied[12,15,39]. However, little is known about the photoelectrochemical stability of carbon nitrides.

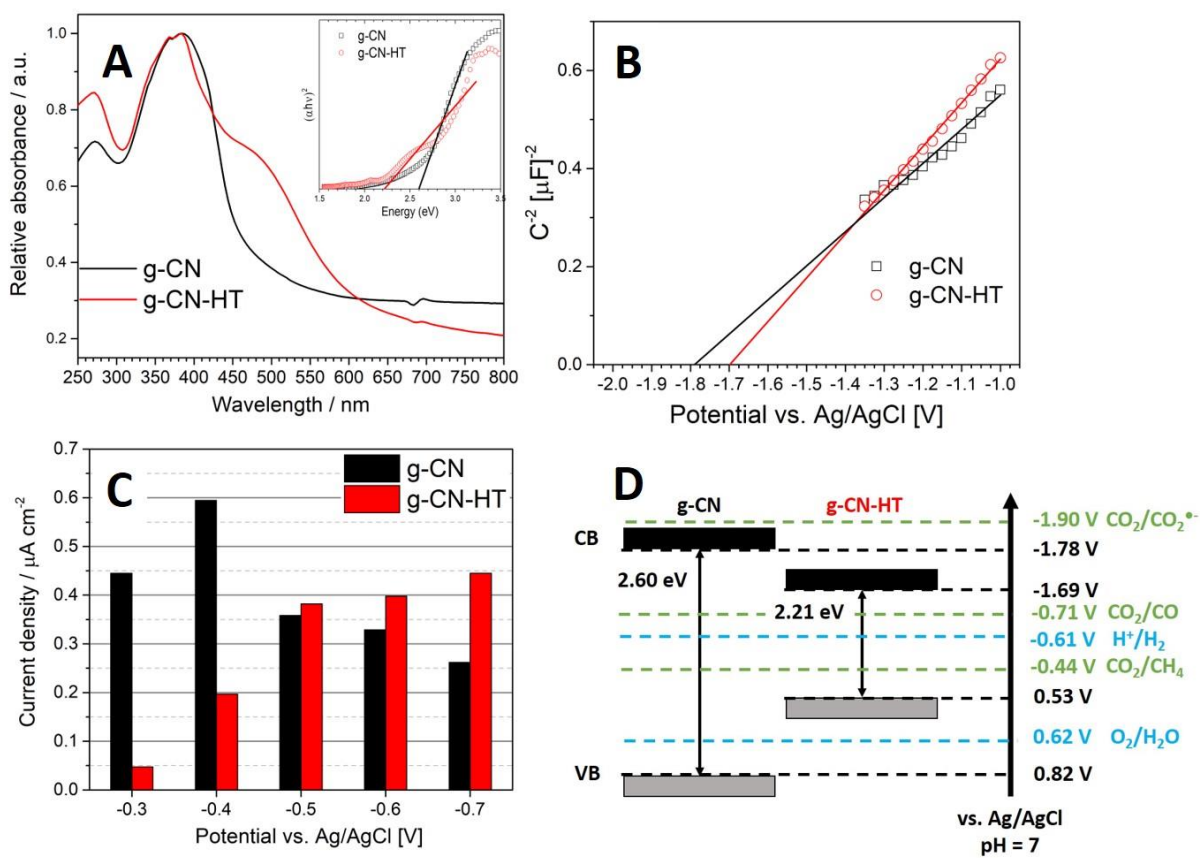

**Figure 3** UV-Vis diffuse reflectance spectroscopy with the Tauc plot in the inset (A); Mott-Schottky plot obtained from dynamic electrochemical impedance spectroscopy at a 100 kHz frequency and 10 mV amplitude (B); photocurrent density determined from chronoamperometry experiments under illumination in $CO_2$ saturated 0.5 M $KHCO_3$ (C) and electronic band structure: grey band represents the valence band; black band represents the conduction band (D)

*Post-mortem analysis*

Different characterization techniques can be used to conduct post-mortem analysis[40]. In fact, the methods are limited only to the nature of the material under investigation. Degradation mechanisms in carbon materials are mostly related to carbon corrosion, such as changes in morphology and surface chemistry. It is well-known that carbon electrodes are prone to corrosion at high anodic potentials[33,41,42]. Carbon oxidation is challenging to assess due to similar values for the wavenumber and binding energy in FT-IR and XPS, respectively. Electron microscopy techniques can be used to track changes in morphology but has limited use in the surface chemistry analysis.

Figure 4 shows the SEM investigations of spray coated g-CN and g-CN-HT on FTO glass before (Figure 4A and C) and after the photoelectrochemical experiments (Figure 4B and D). A uniform coating of the support electrode (FTO glass in this case) with g-CN and g-CN-HT should be achieved to eliminate the exposure of FTO to the electrolyte. Initial g-CN photoelectrodes contain grains which can be as large as 10 µm (Figure 4A) which does not result in a uniform coating. The g-CN-HT photoelectrodes were found to be more uniform (Figure 4C) than g-CN photoelectrodes. The large grains in Figure 4A are the result of g-CN's weak solubility in solvents that lead to poor dispersions. We made use of the spray coating disadvantages to study the photoelectrochemical degradation of g-CN micro grains.

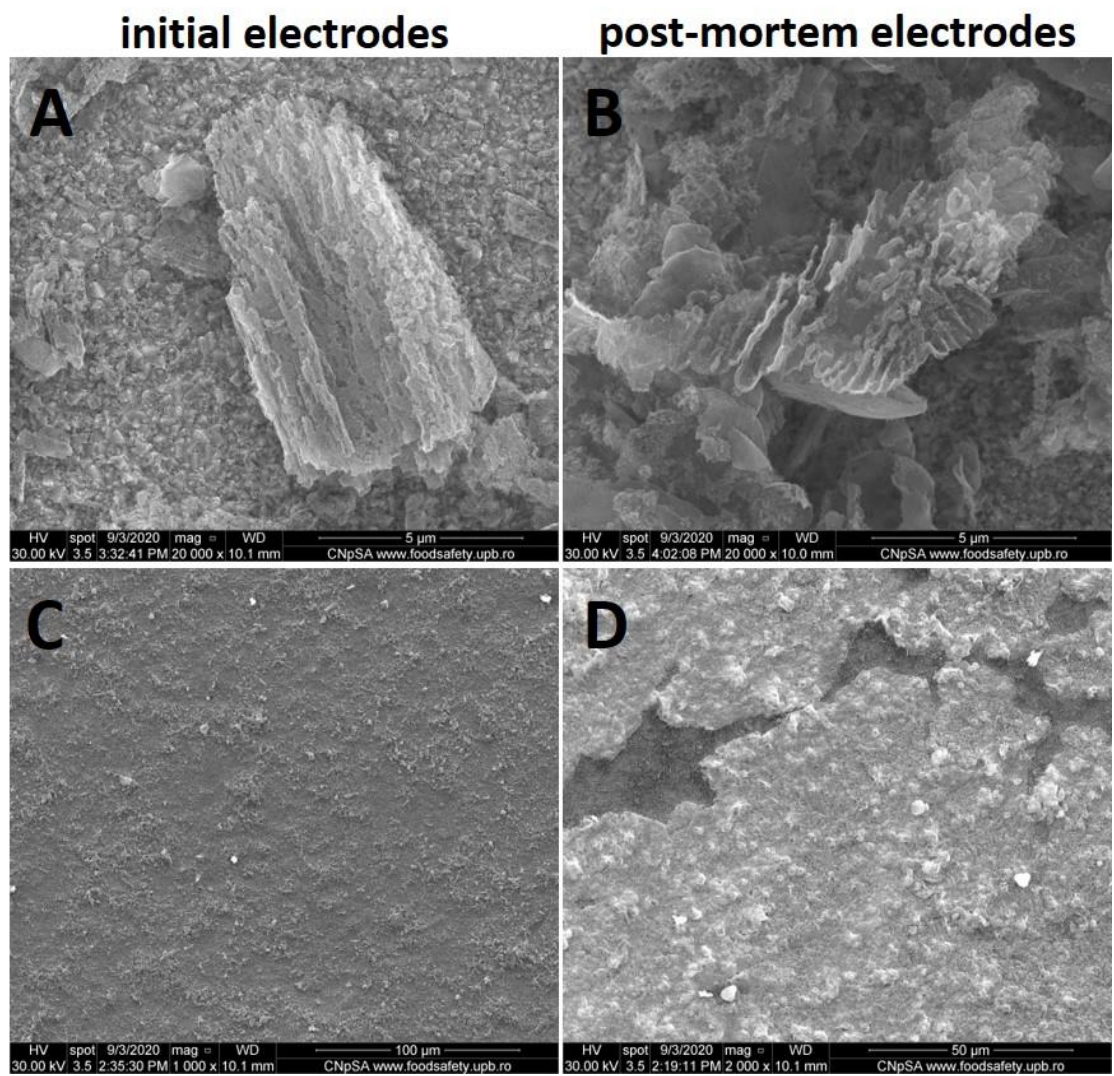

**Figure 4** Scanning electron microscopy of initial and post mortem photoelectrodes with g-CN (A-B) and g-CN-HT (C-D). Initial photoelectrodes are shown in (A and C) and post-mortem photoelectrode are shown in (B and D).

The large grain in Figure 4A reveals that the aligned tube-like structure was preserved during the spray coating. Figure 4B shows the SEM post-mortem investigations, that is the same electrode after the photoelectrochemical experiments. The tube structure of the g-CN is severely altered after only 4 light/dark cycles (Figure 4B). Grains similar to the one in Figure 4A were not found after the photoelectrochemical experiments. Most of the investigated grains were found to be disordered, similar to an amorphous structure. A grain with the thickness of half the tube showing different stages of g-CN degradation is presented in Figure 4B. In the initial degradation stages the tubes are losing their structure along the perpendicular axis of the grain which is in line with the perpendicular mechanism of electron transport in polymeric carbon nitrides[43]. The rest of the grain in Figure 4B is completely amorphous with a structure similar to the heat treated g-CN-HT (Figure S2, Supplementary Material). The obvious degradation can be the root cause for the sample to show a decreasing photocurrent density with increasing potential after -0.4 V. One might argue that the rest of the grain presented in Figure 4B fell in the solution. It is challenging to assume that a grain formed of tightly bound bundle of tubes can be sectioned in such a way by convection only. It is clear that a (photo)electrochemical mechanism is altering the morphology and structure of the initial material.

The initial photoelectrodes containing g-CN-HT (Figure 4C) were considerably smoother than the one with g-CN which may explain the constant increase in the photocurrent density with increasing potential (Figure 3C). g-CN-HT has an amorphous structure (Figure S2, Supplementary Material) which makes it more soluble in solvents but also it makes it more difficult to assess the morphology by SEM, especially when it is deposited in a thin-film. Nevertheless, the deposited film showed pitting behavior in the post-mortem analysis (Figure 4D). Pitting refers to the creation of pits in the films which can be viewed as the darker areas in Figure 4D. To emphasize the lack of pits in the initial sample, the scale bar in Figure 4C was set to 100 µm which is double than the one in Figure 4D. There is no evidence of pitting in the initial sample. Such degradation leads to a poorer photocurrent density over time. Moreover, the film turned brighter in the post-mortem sample (Figure 4D) which is caused by charging. The root cause for charging is the poor electron conductivity between the g-CN-HT, FTO and SEM substrate or the degradation of g-CN-HT that makes it less conductive.

Post-mortem investigations carried out in this work showed that carbon nitride photoelectrodes degrade after a few cycles. It is well-known that the superoxide radical is the first electron transfer in oxygen reduction reaction at nitrogenated carbon electrodes[19,20]. Similarly, $CO_2$ needs to be activated through the 1 electron reaction to the $CO_2$ anion radical[18]. The generation of hydroxyl radical is expected as well at neutral pH under illuminating conditions. Oxygen traces in the electrolyte can lead to the generation of superoxide radical. An XPS study on the stability of polymeric carbon nitrides for photocatalytic $CO_2$ reduction showed that the nitrogen concentration was decreasing while the oxygen concentration was increasing[16]. A decreasing concentration of nitrogen can take place: (1) indirectly, when an increasing concentration of oxygen leads to an apparent decrease in nitrogen or (2) directly, when the carbon nitride decomposing into smaller fractions containing nitrogen that are water soluble.

The electrolyte has been analyzed by UV-Vis to have a better understanding on the degradation of polymeric carbon nitrides. The UV-Vis spectrum of the electrolyte (Figure S5, Supplementary Material) shows two types of degradation for the photoelectrocatalytic experiments (sample biased at -0.7 V vs. Ag/AgCl). A spectrum of the electrolyte during photocatalytic experiments (without applying potential) is shown as a baseline (Figure S5, Supplementary Material) and has one main feature below 250 nm. The region between 300 and 500 nm is specific for the absorption of FTO and carbon nitride which is visible only when the sample is biased, that is photoelectrocatalytic experiments. The other feature is attributed to the reduction of $CO_2$ to formic acid or other similar products (i.e. at 220-230 nm).

Only ring opening can explain the direct loss of nitrogen which has been previously observed in the case of the oxygen reduction reaction at nitrogen doped carbons[44]. Triazine is considered the polymeric carbon nitride building block. Triazine based herbicides are well-known and radical photolysis is one of the degradation approaches to reduce their environmental impact[45–47]. It is generally accepted that hydroxyl radical attack on the triazine unit leads to the generation of nitrate[47]. Nitrates have two characteristic absorption maxima, at 203 and 302 nm, which can be used for quantification. The latter peak is in the region specific for FTO and g-CN absorption (Figure S5, Supplementary Material). The second derivative UV-Vis method[48,49] can be used to determine nitrates down to ppb level. Such a method was used for the electrolyte after 1 h long-term exposure (Figure S6, Supplementary Material). No nitrates were detected which makes carbon nitride photoelectrochemical degradation elusive. Fragmentation of the carbon nitride structure is another probable degradation which needs further investigations, such as liquid chromatography – mass spectrometry or similar. If such a degradation behavior is confirmed by other groups, then the use of polymeric carbon nitrides in photoelectrochemistry needs to be reconsidered.

**Conclusion**

Two different polymeric carbon nitrides were synthesized by thermal polycondensation. A fundamental understanding on the structure and surface chemistry was achieved based on the synchrotron radiation characterization, X-ray diffraction and advanced electron microscopy. The electron band structure was determined by Tauc and Mott-Schottky plots. The product of thermal polycondensation is not the generally accepted graphitic carbon nitride nor melon, although the product is structurally closer to the latter. The materials showed irregular photocurrent increase during the photoelectrochemical $CO_2$ reduction experiments. Post-mortem electron microscopy showed that two degradation mechanisms are at play: (1) nanoscale degradation that alters the initial morphology and most probably the surface chemistry and (2) photoelectrode film structure in which the polymeric carbon nitride film is exfoliating. All in all, the work at-hand shows that polymeric carbon nitrides are prone to degradation, albeit the degradation mechanism is not completely understood. Nevertheless, the extensive use of polymeric carbon nitrides, especially doped carbon nitrides, in photoelectrocatalysis should be carefully reconsidered.

**Author contributions:** FIM and ET have contributed equally. Conceptualization, supervision and methodology SNS; data curation FIM, ET and SNS; formal analysis FIM, BM and CD; investigation TS, LCT, EV and AC; writing-original draft FIM, ET, AC and SNS; resources SA, AC and SNS; writing-review and editing SNS.

# References

1. Fujishima, A. & Honda, K. Electrochemical photolysis of water at a semiconductor electrode. *Nature* **238**, 37–38 (1972).

2. Kato, H., Asakura, K. & Kudo, A. Highly efficient water splitting into H2 and O2 over lanthanum-doped NaTaO3 photocatalysts with high crystallinity and surface nanostructure. *Journal of the American Chemical Society* **125**, 3082–3089 (2003).

3. Wang, X. *et al.* A metal-free polymeric photocatalyst for hydrogen production from water under visible light. *Nature materials* **8**, 76–80 (2009).

4. Cao, S., Low, J., Yu, J. & Jaroniec, M. Polymeric Photocatalysts Based on Graphitic Carbon Nitride. *Advanced Materials* **27**, 2150–2176 (2015).

5. Bojdys, M. J., Müller, J. O., Antonietti, M. & Thomas, A. Ionothermal synthesis of crystalline, condensed, graphitic carbon nitride. *Chemistry - A European Journal* **14**, 8177–8182 (2008).

6. Fina, F. *et al.* Structural investigation of graphitic carbon nitride via XRD and Neutron Diffraction Structural investigation of graphitic carbon nitride via XRD and Neutron Diffraction. *Chemistry of Materials* **27**, 2612–2618 (2015).

7. Lotsch, B. v. Low molecular-weight carbon nitrides for solar hydrogen evolution. *Journal of the American Chemical Society* **137**, 1064–1072 (2015).

8. Lin, L., Ou, H., Zhang, Y. & Wang, X. Tri-s-triazine-Based Crystalline Graphitic Carbon Nitrides for Highly Efficient Hydrogen Evolution Photocatalysis. *ACS Catalysis* **6**, 3921–3931 (2016).

9. Thomas, A. *et al.* Graphitic carbon nitride materials: variation of structure and morphology and their use as metal-free catalysts. *Journal of Materials Chemistry* **18**, 4893 (2008).

10. Akaike, K., Aoyama, K., Dekubo, S., Onishi, A. & Kanai, K. Characterizing Electronic Structure near the Energy Gap of Graphitic Carbon Nitride Based on Rational Interpretation of Chemical Analysis. *Chemistry of Materials* **30**, 2341–2352 (2018).

11. Lotsch, B. v. *et al.* Unmasking Melon by a Complementary Approach Employing Electron Diffraction, Solid-State NMR Spectroscopy, and Theoretical Calculations—Structural Characterization of a Carbon Nitride Polymer.pdf. *Chemistry Euopean Journal* **13**, 4969–4980 (2007).

12. Ong, W. J., Tan, L. L., Ng, Y. H., Yong, S. T. & Chai, S. P. Graphitic Carbon Nitride (g-C3N4)-Based Photocatalysts for Artificial Photosynthesis and Environmental Remediation: Are We a Step Closer to Achieving Sustainability? *Chemical Reviews* **116**, 7159–7329 (2016).

13. Jiang, L. *et al.* Doping of graphitic carbon nitride for photocatalysis: A reveiw. *Applied Catalysis B: Environmental* **217**, 388–406 (2017).

14. Wang, Y., Wang, X. & Antonietti, M. Polymeric graphitic carbon nitride as a heterogeneous organocatalyst: From photochemistry to multipurpose catalysis to sustainable chemistry. *Angewandte Chemie - International Edition* **51**, 68–89 (2012).

15. Ye, S., Wang, R., Wu, M. Z. & Yuan, Y. P. A review on g-C3N4for photocatalytic water splitting and CO2reduction. *Applied Surface Science* **358**, 15–27 (2015).

16. Pomilla, F. R. *et al.* An Investigation into the Stability of Graphitic C 3 N 4 as a Photocatalyst for CO 2 Reduction. *Journal of Physical Chemistry C* **122**, 28727–28738 (2018).

17. Nitopi, S. *et al.* Progress and Perspectives of Electrochemical CO2 Reduction on Copper in Aqueous Electrolyte. *Chemical Reviews* **119**, 7610–7672 (2019).

18. White, J. L. *et al.* Light-Driven Heterogeneous Reduction of Carbon Dioxide: Photocatalysts and Photoelectrodes. *Chemical Reviews* **115**, 12888–12935 (2015).

19. Behan, J. A. *et al.* Untangling Cooperative Effects of Pyridinic and Graphitic Nitrogen Sites at Metal-Free N-Doped Carbon Electrocatalysts for the Oxygen Reduction Reaction. *Small* **15**, 1–10 (2019).

20. Behan, J. A. *et al.* Electrocatalysis of N-doped carbons in the oxygen reduction reaction as a function of pH: N-sites and scaffold effects. *Carbon* **148**, 224–230 (2019).

21. Yang, Y. *et al.* An Unusual Red Carbon Nitride to Boost the Photoelectrochemical Performance of Wide Bandgap Photoanodes. *Advanced Functional Materials* **28**, 1–10 (2018).

22. Kang, Y. *et al.* An Amorphous Carbon Nitride Photocatalyst with Greatly Extended Visible-Light-Responsive Range for Photocatalytic Hydrogen Generation. *Advanced Materials* **27**, 4572–4577 (2015).

23. Diac, C. *et al.* Electrochemical recycling of platinum group metals from spent catalytic converters. *Metals* **10**, 1–11 (2020).

24. Stamatin, S. N., Hussainova, I., Ivanov, R. & Colavita, P. E. Quantifying Graphitic Edge Exposure in Graphene-Based Materials and Its Role in Oxygen Reduction Reactions. *ACS Catalysis* **6**, 5215–5221 (2016).

25. Komatsu, T. The First Synthesis and Characterization of Cyameluric High Polymers, Macromol. Chem. Phys. **202**, 19–25 (2001)

High Polymers

26. Mohamed, N. A. *et al.* The influences of post-annealing temperatures on fabrication graphitic carbon nitride, (g-C3N4) thin film. *Applied Surface Science* **489**, 92–100 (2019).

27. Zheng, Y. *et al.* Hydrogen evolution by a metal-free electrocatalyst. *Nature Communications* 2–9 (2014) doi:10.1038/ncomms4783.

28. Che, W. *et al.* Fast photoelectron transfer in C -C N plane- heterostructural nanosheets for overall water splitting Fast photoelectron transfer in C ring -C 3 N 4 plane-heterostructural. *Journal of the American Chemical Society* **139**, 3021–3026 (2017).

29. Zhang, J. R. *et al.* Accurate K-edge X-ray photoelectron and absorption spectra of g-C3N4 nanosheets by first-principles simulations and reinterpretations. *Physical Chemistry Chemical Physics* **21**, 22819–22830 (2019).

30. Grazioli, C. *et al.* Spectroscopic Fingerprints of Intermolecular H-Bonding Interactions in Carbon Nitride Model Compounds. *Chemistry - A European Journal* **24**, 14198–14206 (2018).

31. Behan, J. A. *et al.* A Combined Optoelectronic and Electrochemical Study of Nitrogenated Carbon Electrodes. *The Journal of Physical Chemistry C* acs.jpcc.6b10145 (2017) doi:10.1021/acs.jpcc.6b10145.

32. Stamatin, S. N., Speder, J., Dhiman, R., Arenz, M. & Skou, E. M. Electrochemical stability and postmortem studies of Pt/SiC catalysts for polymer electrolyte membrane fuel cells. *ACS Applied Materials and Interfaces* **7**, 6153–6161 (2015).

33. Stamatin, S. N. *et al.* Activity and stability studies of platinized multi-walled carbon nanotubes as fuel cell electrocatalysts. *Applied Catalysis B: Environmental* **162**, 289–299 (2015).

34. Deifallah, M., McMillan, P. F. & Corà, F. Electronic and structural properties of two-dimensional carbon nitride graphenes. *Journal of Physical Chemistry C* **112**, 5447–5453 (2008).

35. Jorge, A. B. *et al.* H2 and O2 evolution from water half-splitting reactions by graphitic carbon nitride materials. *Journal of Physical Chemistry C* **117**, 7178–7185 (2013).

36. Albery, W. J., O'Shea, G. J. & Smith, A. L. Interpretation and use of Mott-Schottky plots at the semiconductor/electrolyte interface. *Journal of the Chemical Society, Faraday Transactions* **92**, 4083 (1996).

37. Wang, K. *et al.* Sulfur-doped g-C3N4 with enhanced photocatalytic CO2-reduction performance. *Applied Catalysis B: Environmental* **176–177**, 44–52 (2015).

38. Sagara, N., Kamimura, S., Tsubota, T. & Ohno, T. Photoelectrochemical CO2 reduction by a p-type boron-doped g-C3N4 electrode under visible light. *Applied Catalysis B: Environmental* **192**, 193–198 (2016).

39. Liu, B. *et al.* Phosphorus-Doped Graphitic Carbon Nitride Nanotubes with Amino-rich Surface for Efficient CO2 Capture, Enhanced Photocatalytic Activity, and Product Selectivity. *ACS Applied Materials and Interfaces* **10**, 4001–4009 (2018).

40. Meier, J. C. *et al.* Stability investigations of electrocatalysts on the nanoscale. *Energy & Environmental Science* **5**, 9319 (2012).

41. Yi, Y. *et al.* Electrochemical Degradation of Multiwall Carbon Nanotubes at High Anodic Potential for Oxygen Evolution in Acidic Media. *ChemElectroChem* **2**, 1929–1937 (2015).

42. Hussainova, I. *et al.* A few-layered graphene on alumina nanofibers for electrochemical energy conversion. *Carbon* **88**, 157–164 (2015).

43. Merschjann, C. *et al.* Complementing Graphenes: 1D Interplanar Charge Transport in Polymeric Graphitic Carbon Nitrides. *Advanced Materials* **27**, 7993–7999 (2015).

44. Kim, H., Lee, K., Woo, S. I. & Jung, Y. On the mechanism of enhanced oxygen reduction reaction in nitrogen-doped graphene nanoribbons. *Physical chemistry chemical physics : PCCP* **13**, 17505–10 (2011).

45. da Silva, G., Bozzelli, J. W. & Asatryan, R. Hydroxyl radical initiated oxidation of s-triazine: Hydrogen abstraction is faster than hydroxyl addition. *Journal of Physical Chemistry A* **113**, 8596–8606 (2009).

46. Pellzzetti, E. *et al.* Photocatalytic Degradation of Atrazine and Other s-Triazine Herbicides. *Environmental Science and Technology* **24**, 1559–1565 (1990).

47. Mededovic, S., Finney, W. C. & Locke, B. R. Aqueous-Phase Mineralization of s- Triazine Using Pulsed Electrical Discharge. *Plasma Environmental Science & Technology* **1**, 82–90 (2007).

48. Suzuki, N. & Kuroda, R. Direct Simultaneous Determination of Nitrate and Nitrite by Ultraviolet Second-derivative Spectrophotometry. *Analyst* **112**, 1077–1079 (1987).

49. Crumpton, W. G., Isenhart, T. M. & Mitchell, P. D. Nitrate and Organic N Analyses with Second-Derivative Spectroscopy. *Limnology and Oceanography* **37**, 907 (1992).

# Supporting Information

## Structure and stability of carbon nitrides: ring opening induced photoelectrochemical degradation

Florentina Iuliana Maxim[a], Eugenia Tanasa[b], Bogdan Mitrea[a], Cornelia Diac[a], Tomas Skala[c], Liviu Cristian Tanase[d], Catalin Ianasi[e], Adrian Ciocanea[f], Stefan Antohe[g], Eugeniu Vasile[b], Eugenia Fagadar-Cosma[e], Serban N. Stamatin[a,g]∗

[a] 3Nano-SAE Research Centre, University of Bucharest, Atomistilor 405, 077125, Magurele, Ilfov, Romania

[b] Department of Oxide Materials and Nanomaterials, Faculty of Applied Chemistry and Material Science, University POLITEHNICA of Bucharest, Bucharest 060042, Romania

[c] Department of Surface and Plasma Science, Charles University, V Holešovičkách 2, 18000 Prague, Czech Republic

[d] National Institute of Materials Physics, Atomistilor 405A, 077125, Magurele, Ilfov, Romania

[e] "Coriolan Drăgulescu" Institute of Chemistry, Mihai Viteazul Ave. 24, 300223, Timisoara, Romania

[f] Hydraulics and Environmental Engineering Department, University POLITEHNICA of Bucharest, 060042 Bucharest, Romania

[g] Faculty of Physics, University of Bucharest, Atomistilor 405, 077125, Magurele, Ilfov, Romania

∗corresponding author telephone no.: +40214574838 and e-mail address: serban@3nanosae.org / serban.stamatin@unibuc.ro

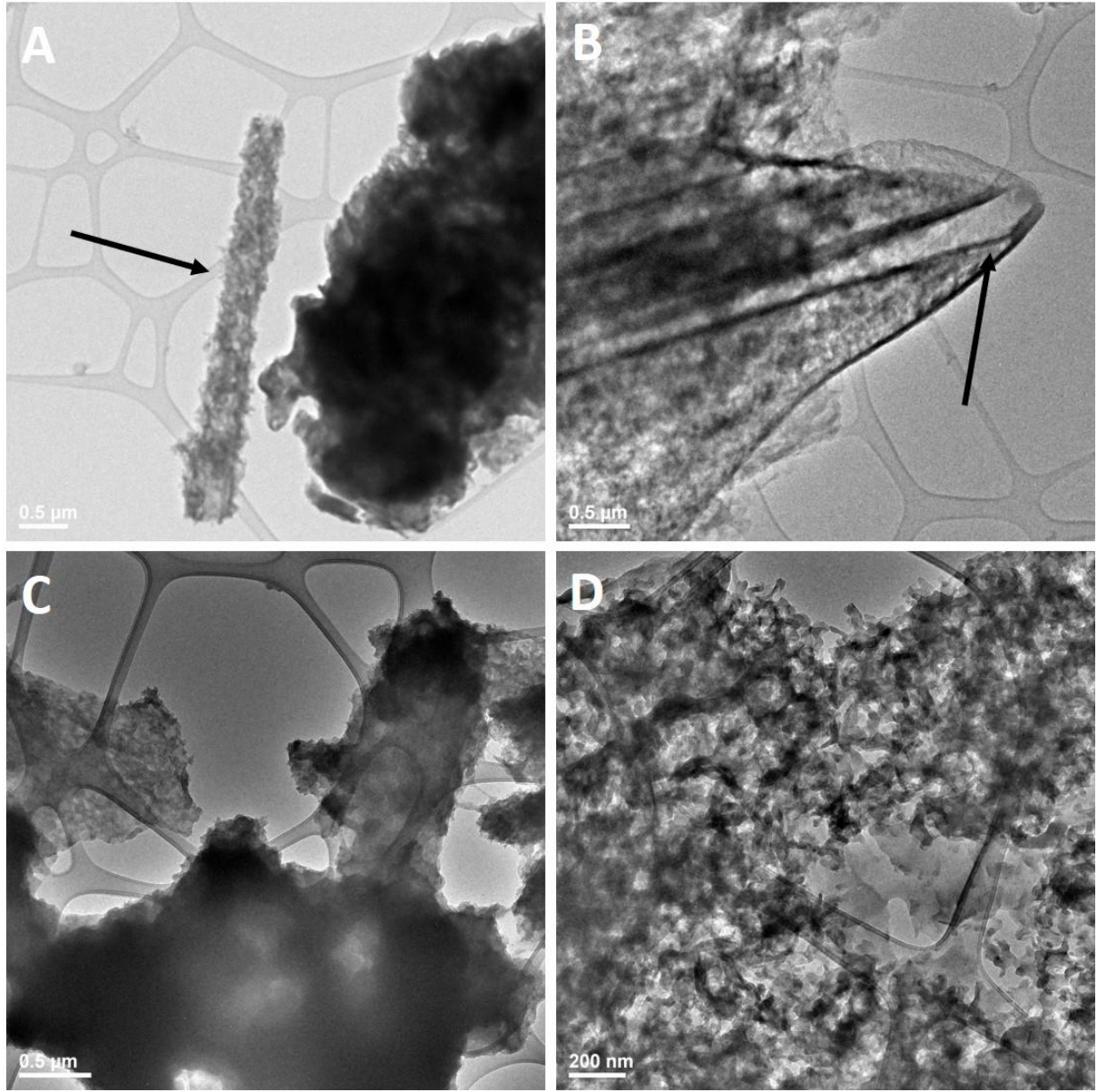

**Figure S 1** Transmission electron microscopy of g-CN (A-B) and g-CN-HT (C-D)

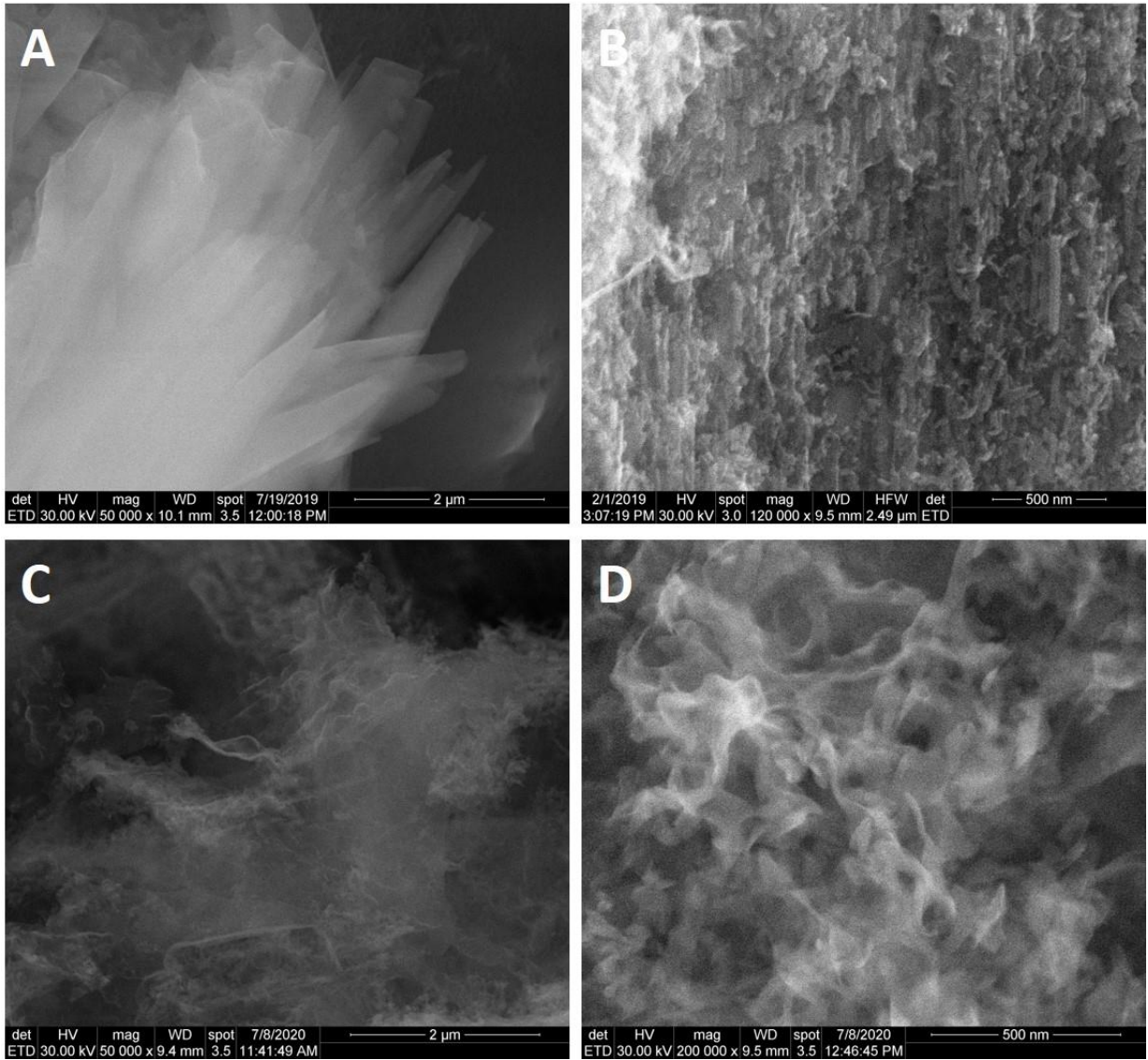

**Figure S 2** Scanning electron microscopy of g-CN (A-B) and g-CN-HT (C-D)

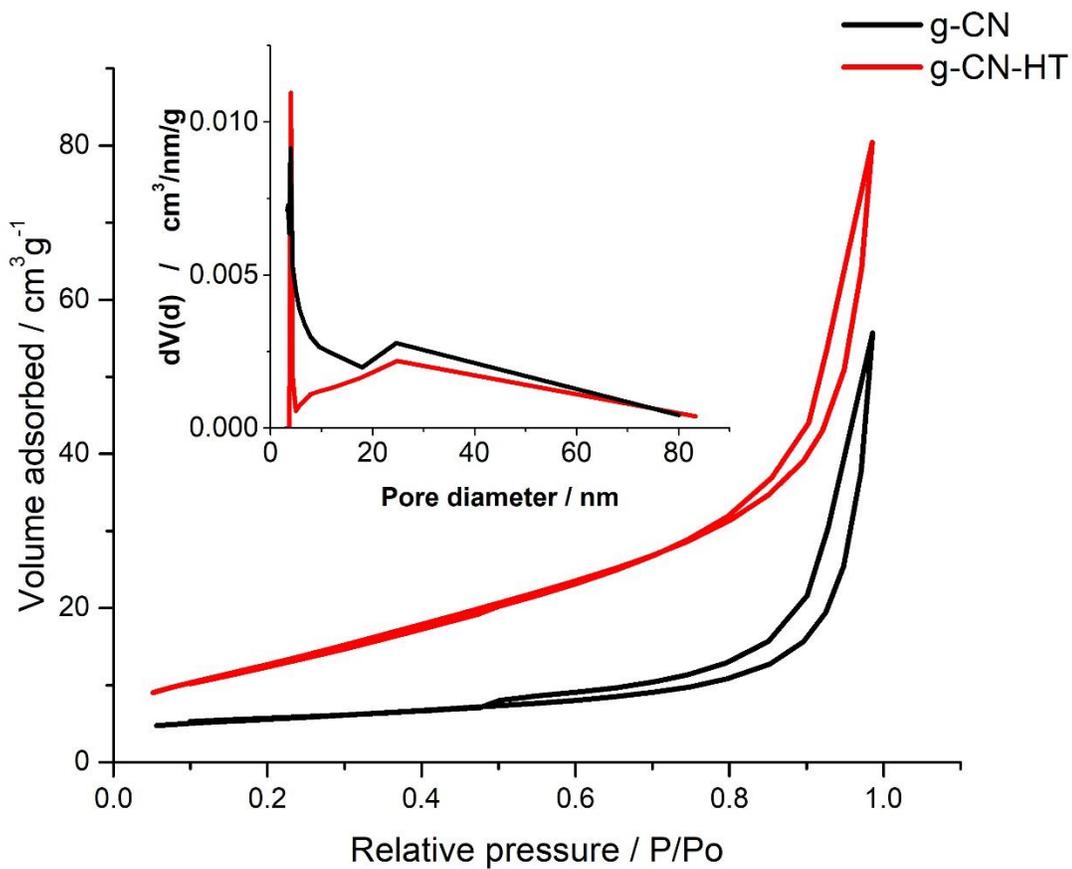

**Figure S 3** $N_2$ adsorption desorption isotherms for samples g-CN(black) and g-CN-HT(red)

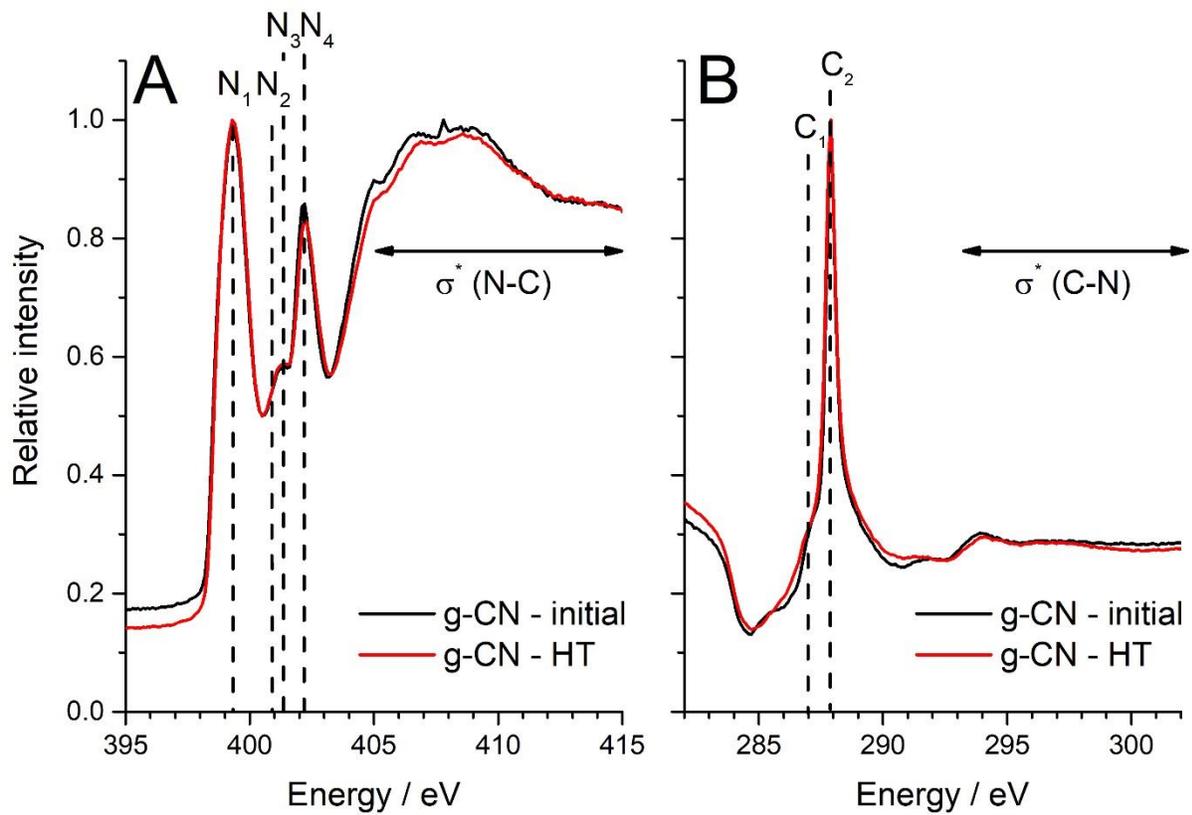

**Figure S 4** Near edge X-ray absorption fine structure of at the N K-edge (A) and C K-edge (B)

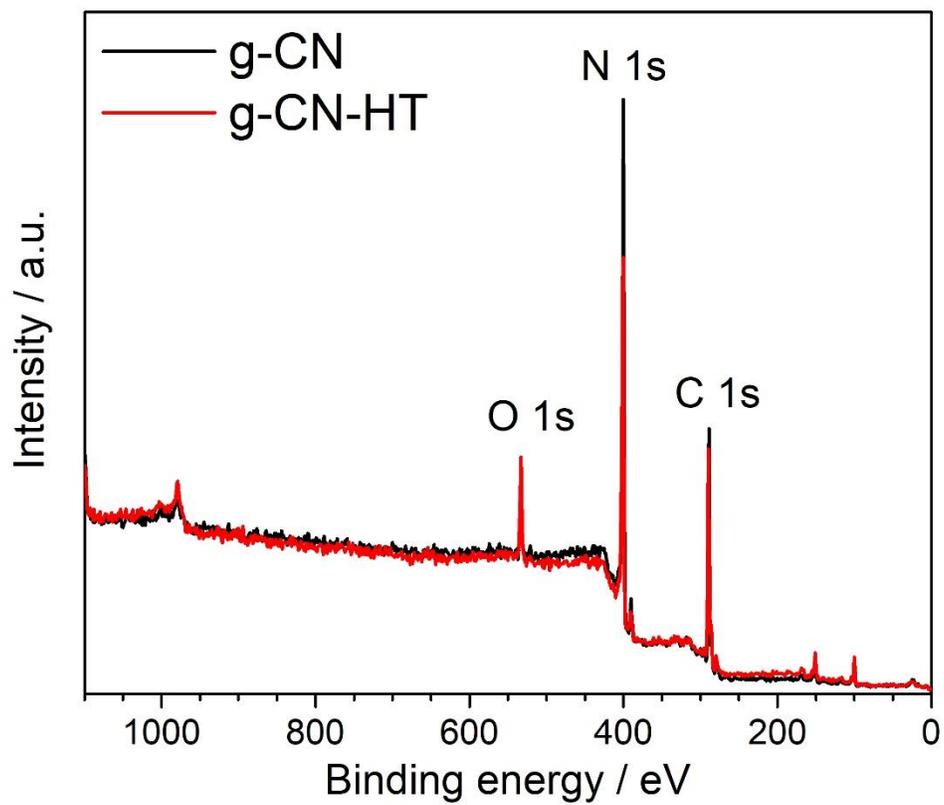

**Figure S 5** Wide spectrum X-ray photoelectron spectroscopy

Mott-Schottky equation:

$$\frac{1}{C^2} = \frac{2}{\varepsilon\varepsilon_0 N_d}\left[(V - V_{fb}) - \frac{k_B T}{e}\right]$$

, where C is the specific capacitance (F cm$^{-2}$), e is the electron charge (=1.602 x 10$^{-19}$ C), $\varepsilon$ is the dielectric constant (i.e. 9.58 for carbon nitride[37]), $\varepsilon_0$ is the vacuum permittivity (=8.854 x 10$^{-14}$ F cm$^{-1}$), $N_d$ is the carrier density, V is the applied potential, $V_{fb}$ is the flat band potential and $k_B T$ is the well-known temperature dependent Boltzmann factor. The value of $k_B T/e$ at room temperature is approx. 0.025 V which is considerably smaller than -1.78 and -1.69 V for g-CN and g-CN-HT, respectively.

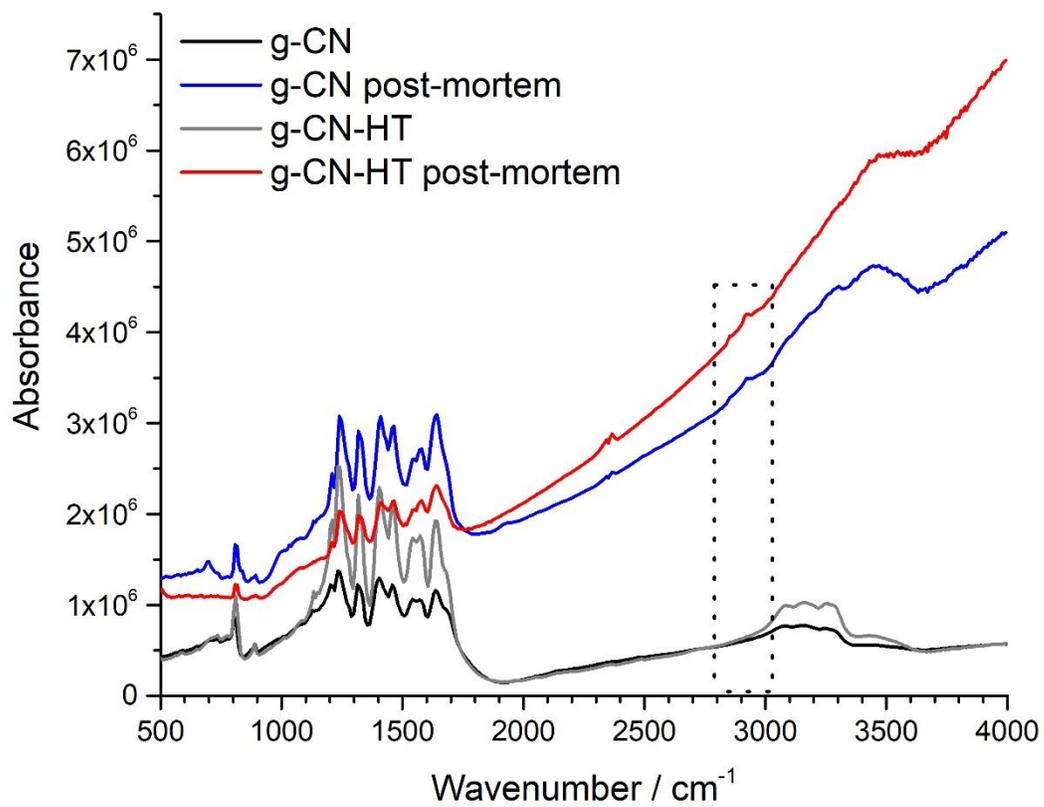

**Figure S 6** Raw FT-IR data

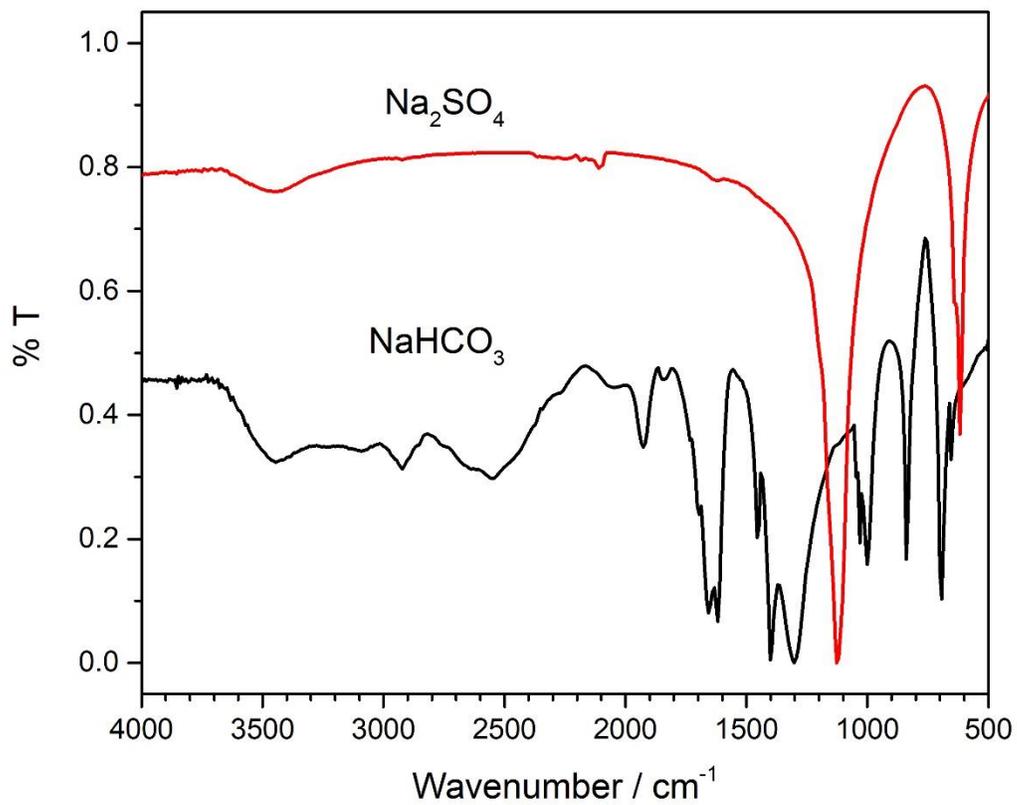

**Figure S 7** FT-IR relative transmittance of $Na_2SO_4$ and $NaHCO_3$

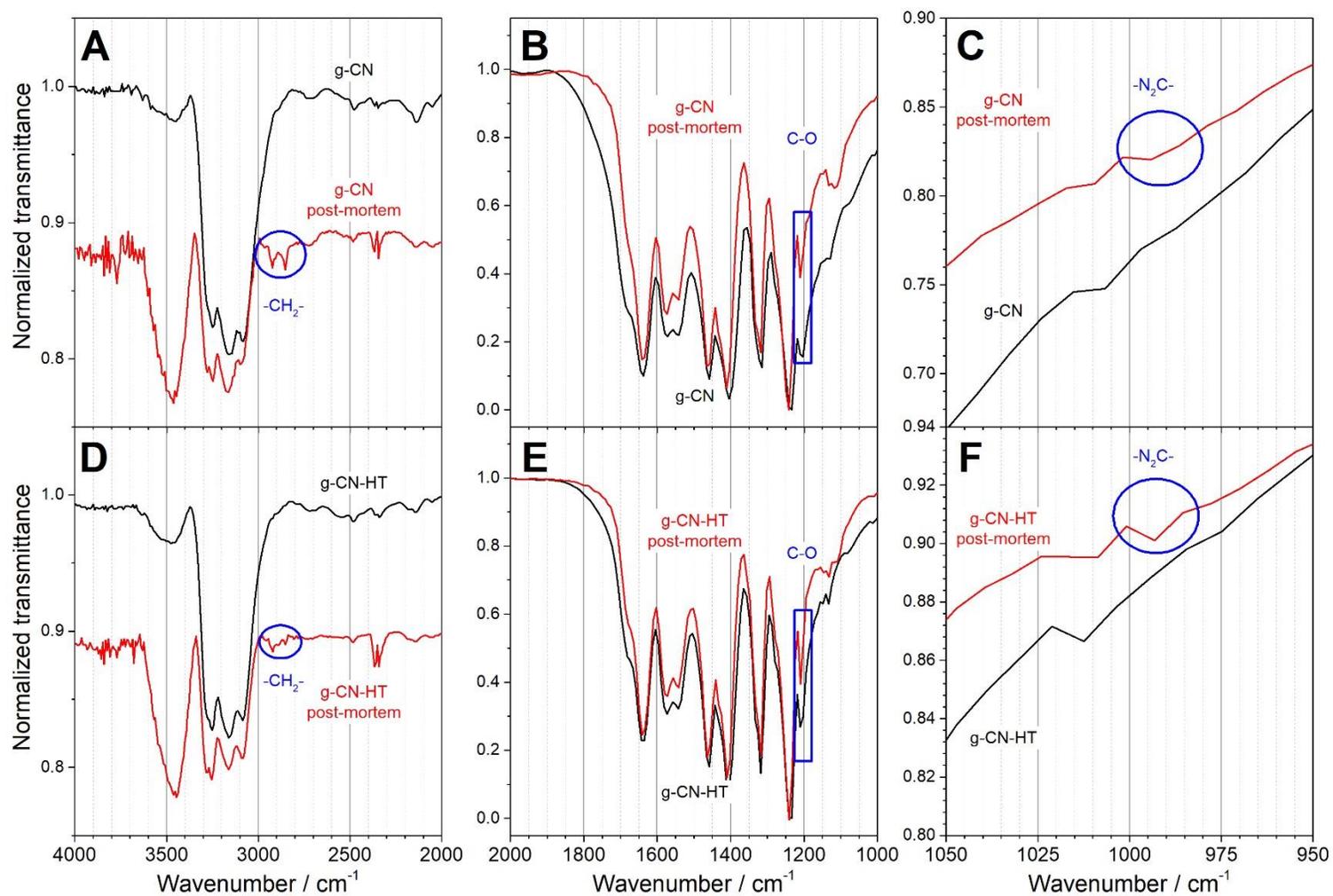

**Figure S 8** FT-IR relative transmittance of initial (black curves) and post-mortem (red curves) of g-CN and g-CN-HT. Blue circles shows the CH$_2$ specific vibration at 2853/2922 cm$^{-1}$. The red curves in A, C, D and F are y-offset by -0.1 .